\documentclass[journal]{IEEEtran}

\ifCLASSINFOpdf
\else
   \usepackage[dvips]{graphicx}
\fi
\usepackage{url}

\usepackage{graphicx}

\usepackage{etoolbox}
\usepackage{amsmath,amssymb,amsfonts,amsthm,bm}
\usepackage{euscript}
\usepackage[T1]{fontenc}
\usepackage[utf8]{inputenc}
\usepackage{nimbusserif}
\usepackage{ifpdf}
\usepackage[english]{babel}
\usepackage{color}
\usepackage{comment}
\usepackage[noadjust]{cite}

\newtheoremstyle{remarkstyle} 
    {\topsep}                    
    {\topsep}                    
    {}                   
    {}                           
    {\itshape}                   
    {.}                          
    {.5em}                       
    {}  

\theoremstyle{definition}
\newtheorem{definition}{Definition}[section]

\newtheorem{theorem}{Theorem}

\theoremstyle{remarkstyle}

\usepackage{tikz-cd}

\usetikzlibrary{decorations.markings,arrows.meta,bending}
\tikzset{->-/.style={decoration={markings,
  mark=at position #1 with {\arrow[line width=2pt]{>}}},postaction={decorate}}}

\begin{document}

\title{Topological IIR Filters over Simplicial Topologies via Sheaves}

\author{Georg Essl, \IEEEmembership{Member, IEEE}
\thanks{Manuscript received April 24, 2020; revised June XX, 2020; accepted
June 21, 2020. Date of publication XXXX XX, 20XX; date of current version
XXXX XX, 20XX. This work was supported by a fellowship of the John Simon Guggenheim Memorial Foundation.}
\thanks{The author is with the Department of Mathematical Sciences at the University of Wisconsin - Milwaukee  (e-mail: essl@uwm.edu).}
}
\markboth{Journal of \LaTeX\ Class Files, Vol. 14, No. 8, August 2015}
{Shell \MakeLowercase{\textit{et al.}}: Bare Demo of IEEEtran.cls for IEEE Journals}
\maketitle

\begin{abstract}
Topology offers a means to formally generalize digital filtering methods based on digital linear translation-invariant (LTI) filters while also, in principle, incorporating translation-variant and nonlinear methods as well as studying large scale (global) properties of filter problems. In this letter we show how the full content of LTI digital filter theory can be incorporated into the formalism of topological filters as introduced by Robinson. In particular, we will give the feedback filter construction associated with infinite impulse responses (IIR). The result allows for direct translation of LTI filters into topological filters, which are sheaves of finite vector spaces and suitably constructed linear maps over simplicial topologies.
\end{abstract}


\IEEEpeerreviewmaketitle

\section{Introduction}

\IEEEPARstart{T}{\lowercase{opological}} signal processing has recently been proposed as an approach to formally generalizing digital signal processing techniques by Robinson \cite{robinson2014topological}. He showed that linear translation-invariant (LTI) finite impulse response (FIR) filters can be incorporated into filtering over topological spaces via sheaves \cite{robinson2013understanding,robinson2014topological}. While not providing explicit construction he suggested that infinite impulse-response filters can also be incorporated by extension to infinite-dimensional vector spaces \cite[Remark 3.5]{robinson2014topological} (or in this context equivalently sequence spaces \cite{robinson2013understanding}). Proper embedding of LTI filter theory in topological filter theory allows one to leverage the wealth and maturity of LTI theory in this framework. The goal of this letter is to give the complete construction of standard (pole-zero) LTI digital filters in Robinson's framework and hence complete a constructive proof that LTI filter theory is fully contained in topological filter theory. Robinson's original proof \cite[Proposition 3.4]{robinson2014topological} is only constructive for FIR filters as the IIR part \cite[Remark 3.5]{robinson2014topological} implies an infinite-dimensional convolution, which is well-known not to be practically realizable directly (compare \cite[Section 2.4]{proakis2006digital}). The content of classical LTI IIR filter theory can best be understood as finitely generated linear recurrence systems which can and, in typical cases, do exhibit infinite impulse responses. As a corollary of our construction we will prove that {\em finite} dimensional vector spaces suffice for realizing all classical LTI filters as topological filters. Our steps are constructive and so can be used to translate any LTI filters into sheaf-theoretical topological filter structures hence allowing a wealth of existing digital filters to be practically realized in this framework. 

\section{Motivation and Related Work}

While the development of topology is predominantly in the realm of pure mathematics \cite{munkres2000topology}, topological ideas have increasingly been found useful in applied domains over the last two decades through the emergence of the fields of computational topology \cite{edelsbrunner2010computational}
, applied topology \cite{ghrist2014elementary}, and most recently the direct application of topological constructions and ideas to data analysis \cite{carlsson2009topology} 
and signal processing \cite{robinson2014topological}.  Furthermore, graph signal processing \cite{Ortega2018GraphSignalProcessing} can be understood as signal processing over low-dimensional simplicial complexes.

Of particular relevance is recent work generalizing linear-translation invariant FIR filters \cite{robinson2013understanding,robinson2014topological} and linear recurrence equations \cite{ghrist2014elementary} using {\em sheaves}. The basic intuition of a sheaf is the ability to connect data locally to a topological space, while retaining consistency of data while traversing the topological space, and the data attached to it. This in turn allows the computation of global solutions. This approach is attractive because it directly tackles the relationship between topological structure and data. While topological ideas have variably been explored in relation to signal processing, perhaps most frequently in the form of time-delay embeddings \cite{takens1981detecting,perea2015sliding,robinson2016quasiperiodic} or other special purpose constructions such as injecting M\"obius band orientability into digital waveguide filters \cite{trautmann1995physical} or using loop spaces to study interaction patterns of digital waveguide filters \cite{Essl:2005ph}, a sheaf-theoretic approach suggests a general strategy in which topology can systematically be used to advance solutions and alternative paths to understanding signal processing problems.

\section{Basic Topology of low-dimensinal simplices}

A background Euclidean space is usually implied in traditional signal processing constructions. In broad areas of application it is customary to imagine the input and output signals to be a continuous function that has been sampled along one dimension. The abstraction of the topological approach can be understood as moving away from the rigid plane and making one dimension (the base space of the signal) a suitable topological space.

A convenient class of topological spaces in this setting are constructed from abstract simplices \cite{edelsbrunner2010computational}, which can be fruitfully be understood to be topological versions of basic geometric shapes of varied dimensions. Our exposition will only need a topological generalization of the real line with discretization, hence we will only need a notion of a local point, and local $1$-dimensional connectivity. A $0$-simplex can be thought of as a topological point. A 1-simplex is a path-connection terminated by two points ($0$-simplices). An extended line can be constructed by gluing $1$-simplices together at a shared $0$-simplex. If there is no branching we will call this simplicial complex a {\em line complex}.
Observe that this construction looks quite analogous to a sampled version of the real line $\mathbb{R}$, except that there is none of the metric structure of $\mathbb{R}$. Observe that in fact any irregularly sampled real line is an example of the line complex. Hence we say that any sampled real lines share the same topological structure. Based on this observation, we can say that topological signal processing over line complexes contains regular, irregular \cite{marvasti2001nonuniform} and event-based sampling \cite{miskowicz2018event} under addition of the appropriate metric structure.

The following two operations allows one to relate $n$-simplices to each other in a simplicial complex, hence arrive at a way to traverse the complex. For our purposes, the order of simplices, preserved by these maps, can be associated with the direction of traversal.

\begin{definition}
A {\bf boundary operation} $b$ of an $n$-simplex $\mathcal{X}_n$ returns an ordered set of all $\text{n-1}$-simplices that constitute its boundary. A $0$-simplex returns the empty set. A {\bf face operation} $f$ of an $n$-simplex returns an ordered set of all $n+1$-simplices of which it is a boundary. We call two simplices {\bf directly connected} if their their are relatable through one face or boundary operation.
\end{definition}

\section{Associating Data to a Topology via Sheaves}

Next, we need to attach data to the topological space. Sheaf theory was developed starting in the 1940 for this very purpose \cite{gray1979fragments} and became an important building block in algebraic topology and algebraic geometry. The construction of attaching data to cellular or simplicial structures emerged later \cite{shepard1986cellular,curry2014sheaves} with applications emerging only recently \cite{ghrist2014elementary,robinson2014topological}.

The general definition of sheaves is designed to work over general topological spaces \cite{bredon1997sheaf}, providing an excessive level of abstraction for the purpose of this letter. Furthermore, computation of (co)homology is often a central goal. This, too, is not a direct goal of this paper, though sheaf (co)homology can be fruitfully used for a range of interesting derivations \cite{robinson2014topological,ghrist2014elementary}. Hence we will use the following definition of a sheaf:

\begin{definition}
A {\bf sheaf} $\mathcal{S}$ of data $\mathcal{D}$ of a simplicial complex $\mathcal{X}$ each indexed by $i\in\mathcal{I}$ satisfying
\begin{enumerate}
    \item Each data $\mathcal{D}_i$ is attached to each $n$-simplex $\mathcal{X}_i$.
    \item Local data $\mathcal{D}_i$ is unique. We call this the {\bf consistency condition}.
    \item If two simplices $\mathcal{X}_i,\mathcal{X}_j$ are {\em directly connected}, then there exists at least one mapping between $\mathcal{D}_i$ and $\mathcal{D}_j$. We call such a mapping the {\bf consistency map}\footnote{Traditionally, this map is called {\em restriction map} in classical sheaf theory. This name is justified because all possible data is restricted to achieve local-to-global data consistency. Given that restriction is not apparent in our construction, but consistency is, we opt for the more descriptive name.}.
\end{enumerate}
\end{definition}

\noindent While not quite as general as sheaves over arbitrary topological spaces, this is still a very general definition from a signal processing perspective. Note that any simplicial complex is permissible as is the nature of the data. The main difference of our definition to other proposals \cite{ghrist2014elementary,robinson2014topological} is the emphasis of local consistency, in contrast to applications that look to study ambiguity through sheaves (as is done in \cite{robinson2013understanding}). Imposing local consistency on all our constructions retains the classical behavior implied in traditional digital filters. Furthermore, we avoid overly explicit use of category theory, not because of its complexity in this situation, but because it introduces another layer of language not necessary for this exposition\footnote{In terms of category theoretical language sheaves are typically defined by attaching data via functors, in our case between a map defined over the topological space such as the boundary operator and a map on the data that is attached.}.
We will use arrows that the reader unfamiliar with category theory can read as simply mappings between spaces of data.

\section{Topological Filters via Sheaves}\label{sec:topfilt}

The general definition of a {\em topological filter} via sheaves after Robinson \cite[Definition 4.15]{robinson2014topological} is:
\begin{align}\label{eq:topfilt}
    \mathcal{S}_i\xleftarrow{i}\mathcal{S}_{s}\xrightarrow{o}\mathcal{S}_o
\end{align}
The sheaves in this construction have the following meaning: $\mathcal{S}_i$ is the input sheaf, $\mathcal{S}_o$ is the output sheaf, and $\mathcal{S}_{s}$ is the state of the topological filter. This general structure is broadly applicable to all sorts of data, and the nature of the input and output maps is general. For the direction of the arrows, we follow Robinson's convention \cite{robinson2014topological}. We call these sheaves because we think of this structure being attached over simplices of our topological space and connected via consistency maps. Hence, tracing out an underlying simplicial line complex we get a general structure as follows:

\begin{figure}[ht]
\centering
\begin{tikzcd}
\cdots \arrow{r} & 0 &\arrow{l} \mathcal{S}_i \arrow{r} & 0 & \arrow{l} \cdots\\
\cdots \arrow{r} & \mathcal{S}_{c} \arrow{u} \arrow{d}  &\arrow{l}{r} \mathcal{S}_{s} \arrow{u}{i} \arrow{d}{o} \arrow{r}{s} & \mathcal{S}_{c} \arrow{u} \arrow{d} & \arrow{l}  \cdots\\
\cdots \arrow{r} & 0 &\arrow{l} \mathcal{S}_o \arrow{r} & 0 & \arrow{l} \cdots
\end{tikzcd}
$\cdots$\includegraphics[width=0.7\columnwidth]{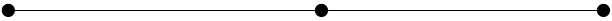}$\cdots$
\caption{The sheaf filter structure over a line complex. The vertical sheaf structure is associated with the $0$ and $1$-simplex below it. Vertical maps relate to traversal between neighboring simplices via face and boundary maps.\label{fig:sheafline}}
\end{figure}

\noindent Note that the vertical slices of Figure \ref{fig:sheafline} have the basic form of the general topological filter of equation (\ref{eq:topfilt}). The state over each $0$-simplex $\mathcal{S}_{s}$ is connected via a consistent sheaf $\mathcal{S}_{c}$ over the connecting $1$-simplices. These capture the information that needs to be retained to keep state transitions consistent with the desired filter computation as the filter state is updated. The top row corresponds to inputs over different simplices of the underlying topological space. The bottom row corresponds to the output at the same simplices. Given that input and output samples are treated as independent -- there is no information to consider for their respective consistency in transition -- we have $0$ sheaves over $1$-simplices for each \cite{robinson2014topological}. Maps in this diagram are called {\em sheaf maps}.

In principle all $\mathcal{S}_*$ can be very general mathematical objects: algebraic, vector, geometric, function or topological spaces and suitable mappings defined over them. Linear systems theory and linear digital signal processing operate over finite vector spaces, with linear or affine maps between them.

\section{Digital FIR Filters as Sheaves}\label{sec:FIR}

One important basic class of LTI filter structures are finite impulse response (FIR) filters \cite{proakis2006digital}.
This construction is due to Robinson \cite{robinson2013understanding,robinson2014topological}. We follow his exposition to explain the basic properties of digital filters over sheaves and to illustrate the relationship of this construction to the IIR construction we will derive in section \ref{sec:iir} as well as to allow us to use this result to construct the general pole-zero filter on sheaves.

An FIR filter is a weighted sum of delayed versions of a signal over a given length $N$. We denote the weights as $b_i$ (which in this case also correspond to the impulse response), the current input signal as $x$, and the output as $y$ and the past state vector $x_i$. The resulting response of the filter is:
\begin{align}
y&=\sum_{i=1}^{N}{b_i\cdot x_{N-i}}+b_0\cdot x\label{eq:fireq}
\end{align}
A state of length $N+1$ is retained to implement this computation, hence suggesting that standard digital FIR filters can be rewritten as sheaves as follows \cite{robinson2013understanding,robinson2014topological}:
\begin{figure}[ht]
\centering
\begin{tikzcd}
\cdots \arrow{r} & 0 &\arrow{l} \mathbb{R} \arrow{r} & 0 & \arrow{l} \cdots\\
\cdots \arrow{r} & \mathbb{R}^{N} \arrow{u} \arrow{d}  &\arrow{l}{r} \mathbb{R}^{N+1} \arrow{u}{i} \arrow{d}{o} \arrow{r}{s} & \mathbb{R}^{N} \arrow{u} \arrow{d} & \arrow{l}  \cdots\\
\cdots \arrow{r} & 0 &\arrow{l} \mathbb{R} \arrow{r} & 0 & \arrow{l} \cdots
\end{tikzcd}
\caption{FIR filter in sheaf form using linear maps between vector spaces \cite{robinson2014topological,robinson2013understanding}. This structure turns out to be identical in terms of vector space dimensionality the IIR case.\label{fig:sheaffir}}
\end{figure}

\noindent The center row in Figure \ref{fig:sheaffir} corresponds to the state of the filter. In this diagram we have now specified that we operate on finite vector spaces.
The linear map $i$ injects the input and $o$ computes the output. The left column represents the transition from an earlier state where information is {\em consistently} contributed to the next state via ($r$). 
The right column represents the shifting ($s$) of the information for the next time step. Give that all maps are linear maps between vector spaces, they can be realized as matrix multiplications (or simplifications thereof).
Observe that the consistency sheaf $\mathcal{S}_c$ actually contains one less state dimension than the full state $\mathcal{S}_s$ during output computation. This captures the fact that during temporal update, one element of the state vector is shifted outside the bounds of the state and hence seizes to contribute to a future state. In a more sheaf-theoretical language, that part of the old state is not required for consistency with the new state. 

\noindent We can complete the definition of all sheaf maps from equation \ref{eq:fireq} and from the shift of the filter state \cite{robinson2014topological}:
\begin{align}
s&: (x_0,x_1,...,x_{N-1},x) \rightarrow (x_1,x_2,...,x_{N-1},x)\label{eq:firshift}\\
r&: (x_0,x_1,...,x_{N-1},x) \rightarrow (x_0,...,x_{N-1})\label{eq:firr}\\
i&: (x_0,...,x_{N-1},x) \rightarrow (x)\label{eq:firinput}\\
o&: (x_0,x_1,...,x_{N-1},x) \rightarrow (b_0\cdot x+\sum_{i=1}^N{b_i\cdot x_{N-i}})\label{eq:firoutput}
\end{align}
The linear map $s$ has the form of a shift matrix.
As a practical matter, maps $r$ and $i$ are computed in practical implementation in the opposite direction of the arrow. With this in mind note that that all computations are completely identical to those of classical LTI FIR filters.
\section{Digital IIR Filters as Sheaves}\label{sec:iir}

The general IIR filter admits both feedback and feedforward computation. However, the crucial aspect of IIR filters is characterized by the feedback behavior, and general IIR filters can be recovered as a pure feedback (also known as all-pole) IIR filter concatenated with a pure feedforward FIR filter. Hence it provides clarity and simplifies the discussion to only consider all-pole filters first, which we will develop next.

\subsection{All-Pole IIR Filters}

An all-pole filter of length $N$ with current input $x$ and $y$ and past state $y_j$ is defined as:
\begin{align}
    y&=\sum_{j=1}^{N}{-a_j\cdot y_{N-j}}+x
\end{align}
This can be reformulated in terms of linear equations over a state space with $\bm{x}$ being the state vector, $\bm{y}$ the output vector and $\bm{u}$ the input vector, as follows \cite{FILTERS07}:
\begin{align}
    \bm{y}=\bm{C}\cdot \bm{x} + \bm{D}\cdot \bm{u}\\
    \bm{x}=\bm{A}\cdot \bm{x} + \bm{B}\cdot \bm{u}\label{eq:statespace}
\end{align}
The feedback dynamics is captured by the square $N\times N$ matrix $A$ (here we use the canonical controllability form:
\begin{align}
    \bm{A}=
  \begin{bmatrix}
 0      & 1     &  0     & \cdots      & 0 \\
 0      & 0     &  1     & \cdots      & 0 \\
 \vdots & \vdots & \vdots & \ddots &  \vdots \\
 0      &  0     & 0 & \cdots  & 1 \\
-a_N      &  -a_{N-1}     &  -a_{N-2}     & \cdots     &  -a_1 
\end{bmatrix}
\end{align}
The state transition matrix $\bm{A}$ captures the transition, and in this case the feedback of the IIR filter. Notice that the FIR case of transition via a shift matrix can be recovered when all feedback coefficients $a_j$ are set to zero.

In the case of a pure all-pole filter, the remaining linear maps ($\bm{B}, \bm{C}, \bm{D})$ of the state space model of equation (\ref{eq:statespace}) are:
\begin{align*}
\bm{B}&=\begin{bmatrix}0  & \cdots & 0 & 1\end{bmatrix} &
\bm{C}&=\begin{bmatrix}0 & \cdots & 0 & 1\end{bmatrix} &
\bm{D}&=\begin{bmatrix} 1 \end{bmatrix}
\end{align*}
To convert this model to the sheaf structure, we observe that $\bm{A}$ corresponds directly to the composition of maps $s$ and $r$, that is the maps that describe how information is consistently shared between state iterations (this part of the filter dynamic can be understood as linear recurrence, see \cite[example 9.1]{ghrist2014elementary}). $\bm{B}$ and $\bm{D}$ represent how the input $x$ is connected to both the state hence relating to the map $x$, and the output $y$ hence partly relating to the map $y$, respectively. Finally, $\bm{C}$ corresponds to the how the state relates to the output, hence describing the remaining part of $y$.

In order to maintain linear maps throughout the sheaf structure, we have to express all aspects of the state space model in terms of linear maps (which is, given suitable basis vectors equivalent to matrix multiplication). However, the standard state space model is affine (due to the input being additive). We utilize the standard trick of using homogeneous coordinates to convert affine maps into linear maps \cite{gallier2019algebra} by introducing one extra dimension that incorporates translations (vector additions) into the linear map.
The All-Pole IIR filter behavior can then be realized with the following sheaf maps:
\begin{align}
\begin{split}
s: (x_0,x_1,...,x_{N-1},x) \rightarrow&  (x_1,x_2,...,x_{N-1},\\&  x+\sum_{j=1}^{N}{-a_j\cdot x_{N-j}})\label{eq:allpoleiirshift}
\end{split}\\
r: (x_0,x_1,...,x_{N-1},x) \rightarrow&  (x_0,x_1,...,x_{N-1})\\
i: (x_0,x_1,...,x_{N-1},x) \rightarrow&  (x)\\
o: (x_0,x_1,...,x_{N-1},x) \rightarrow&  (x_{N-1}+x)
\end{align}
\subsection{Pole-Zero IIR Filters}
The filter equation for the general IIR filter contains feedforward and feedback contributions with current input $x$, output $y$, and past filter states $x_i$ and $y_j$, respectively, as follows:
\begin{align}
 y&=\sum_{i=1}^N{b_i\cdot x_{N-i}} + b_0\cdot x +\sum_{j=1}^{N}{-a_j\cdot y_{N-j}}\label{eq:geniir}
\end{align}
In this form we write all summations in terms of the higher order filter $N$. This contains filters of arbitrary length differences by setting coefficients to zero as needed.
\begin{figure}[th]
\centering
\includegraphics[width=0.75\columnwidth]{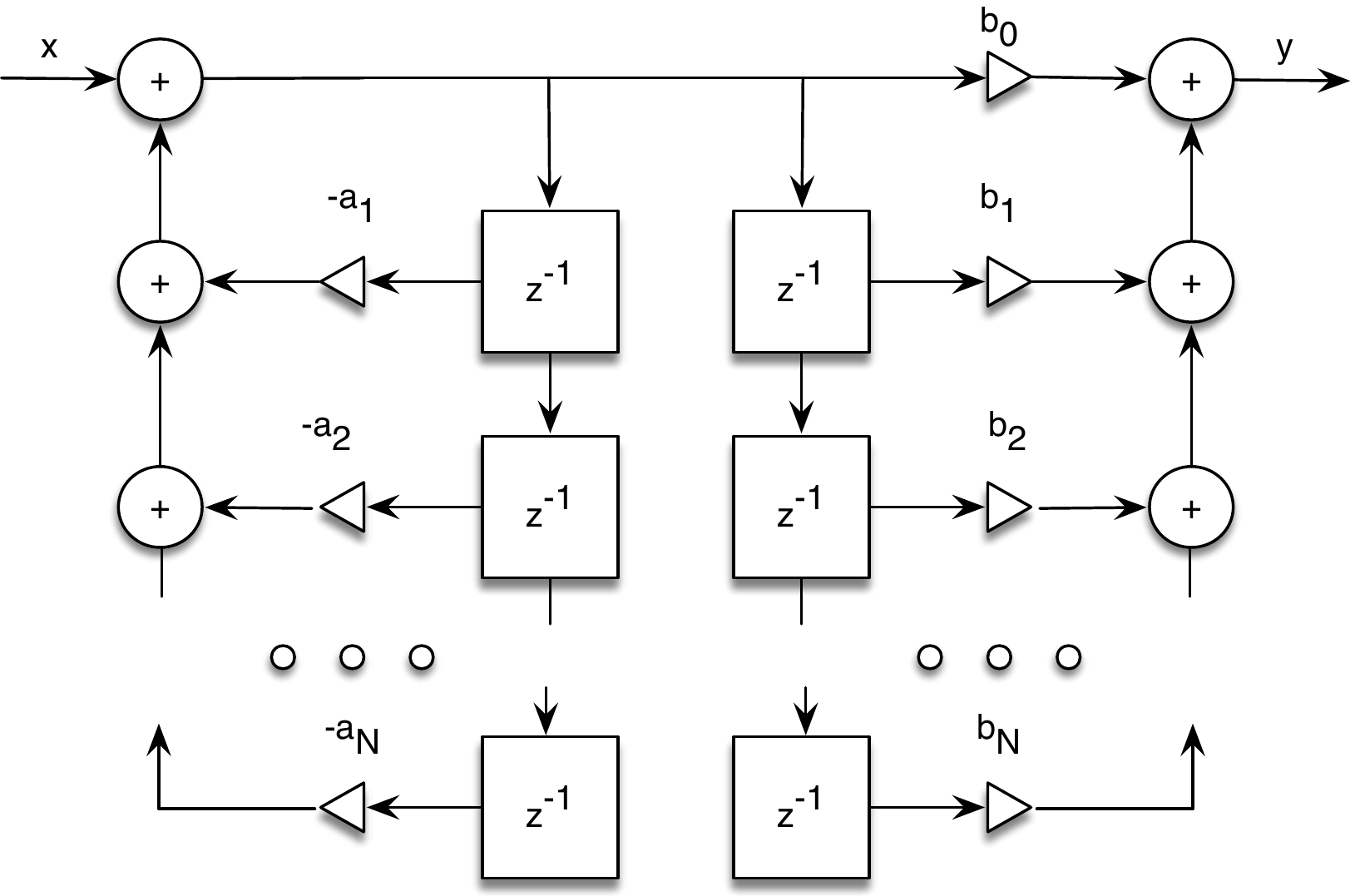}
\caption{General IIR filter in split direct form II depicted as all-pole IIR filter followed by an FIR filter of the same order. The respective states are identical hence can be merged, leading to the canonical direct form II.\label{fig:iird2}}
\end{figure}

\noindent There are a number of practical realizations of such filters. However the Direct Form II is particularly convenient for our construction \cite{FILTERS07}.
When we can concatenate the FIR filter structure 
to the All-pole IIR filter structure, 
and require that the order of the FIR filter not exceed the order of the IIR filter (as in equation \ref{eq:geniir}), we get the split direct form II of the filter (Figure \ref{fig:iird2}). This depiction makes clear that the internal state is shared and hence does not need to be duplicated. In this order, the general IIR filter differs from the all-pole filter merely in the output map $o$, which now contains also the feedforward summation of the FIR filter. Hence we arrive at the maps for the general pole-zero IIR filter:
\begin{align}
\begin{split}
s: (x_0,x_1,...,x_{N-1},x) \rightarrow& (x_1,x_2,...,x_{N-1},\\&  x+\sum_{j=1}^{N}{-a_j\cdot x_{N-j}})
\end{split}\\
r: (x_0,x_1,...,x_{N-1},x) \rightarrow& (x_0,x_1,...,x_{N-1})\\
i: (x_0,x_1,...,x_{N-1},x) \rightarrow& (x)\\
o: (x_0,x_1,...,x_{N-1},x) \rightarrow& (b_0 \cdot x + \sum_{i=1}^N{b_i\cdot x_{N-i}})
\end{align}
\noindent This construction constitutes a proof of the following theorem:

\begin{theorem}
General LTI IIR filters (\ref{eq:geniir}) can be realized as topological filters (\ref{eq:topfilt}) via linear maps between finite dimensional vector spaces. Specifically the state sheaf $S_s$ is finite dimensional.
\end{theorem}

\section{Conclusion}
In this letter, we gave the complete construction of the classical theory of linear translation-invariant digital filters as linear maps between sheaves of finite dimensional vector spaces over simplicial topologies. This constitutes a constructive proof that this class of filters are completely contained in the notion of topological filters and can used without restriction in this emerging theory. The goal of this letter is to contribute to the development of a familiar grounding of classical digital signal processing in the framework of topological filtering.

The given construction of LTI filters using sheaves are computationally identical to traditional linear time invariant filter. However, this is now constructed in relation to some arbitrary space of a simplicial topology. This provides a kind of interpretation of the linear translation-invariant character of the filters.
We observe that given any line complex, independent of further structure, which includes any information about spacing of samples, we can associate linear maps of linear translation invariant filters to a sheaf over it. This construction is then formally valid. Hence concrete sampling using metric spacing is external to this construction, but more generally, this construction is valid even if it may be difficult to identify a metric, or the metric is unknown. For example, consider a moving source that whose oscillatory behavior is modeled by an IIR filter. As long as the IIR filter is updated at constant time intervals, the output will correctly capture the behavior for all topologically connected paths that the source must have taken in the interim. The inverse suggests a topological filter construction to detect moving sources. It should be noted that sheaves themselves can be fruitfully be used to model sampling behavior and generalize standard Nyquist-Shannon-style sampling theorems \cite{robinson2015sheaf}. 

Topological filters promise a systematic study of signal processing over topological spaces, and the study of algebraic topological properties in the context of signal processing \cite{robinson2014topological}. Further, it allows for attaching more general mathematical models than linear maps between vector spaces, while having a unifying principle of local consistency to inform filter construction.

\section{Acknowledgments}
Many thanks to Perry Cook and Bob Adams for helpful discussions and pointers to relevant literature. Furthermore, the thoughtful feedback by three anonymous reviewers has helped improve the final version of this letter.


\bibliographystyle{IEEEtran}
\bibliography{IEEESPL-TopologicalIIRFilters} 

\begin{thebibliography}{10}
\providecommand{\url}[1]{#1}
\csname url@samestyle\endcsname
\providecommand{\newblock}{\relax}
\providecommand{\bibinfo}[2]{#2}
\providecommand{\BIBentrySTDinterwordspacing}{\spaceskip=0pt\relax}
\providecommand{\BIBentryALTinterwordstretchfactor}{4}
\providecommand{\BIBentryALTinterwordspacing}{\spaceskip=\fontdimen2\font plus
\BIBentryALTinterwordstretchfactor\fontdimen3\font minus
  \fontdimen4\font\relax}
\providecommand{\BIBforeignlanguage}[2]{{%
\expandafter\ifx\csname l@#1\endcsname\relax
\typeout{** WARNING: IEEEtran.bst: No hyphenation pattern has been}%
\typeout{** loaded for the language `#1'. Using the pattern for}%
\typeout{** the default language instead.}%
\else
\language=\csname l@#1\endcsname
\fi
#2}}
\providecommand{\BIBdecl}{\relax}
\BIBdecl

\bibitem{robinson2014topological}
M.~Robinson, \emph{Topological signal processing}.\hskip 1em plus 0.5em minus
  0.4em\relax Springer, 2014.

\bibitem{robinson2013understanding}
------, ``Understanding networks and their behaviors using sheaf theory,'' in
  \emph{Global Conference on Signal and Information Processing
  (GlobalSIP)}.\hskip 1em plus 0.5em minus 0.4em\relax IEEE, 2013, pp.
  911--914.

\bibitem{proakis2006digital}
J.~G. Proakis and D.~G. Manolakis, ``Digital signal processing,''
  \emph{Pearson}, 2006.

\bibitem{munkres2000topology}
J.~R. Munkres, \emph{Topology}, 2nd~ed.\hskip 1em plus 0.5em minus 0.4em\relax
  Prentice Hall, 2000.

\bibitem{edelsbrunner2010computational}
H.~Edelsbrunner and J.~Harer, \emph{Computational topology: an
  introduction}.\hskip 1em plus 0.5em minus 0.4em\relax American Mathematical
  Society, 2010.

\bibitem{ghrist2014elementary}
R.~Ghrist, \emph{Elementary applied topology}.\hskip 1em plus 0.5em minus
  0.4em\relax Createspace, 2014.

\bibitem{carlsson2009topology}
G.~Carlsson, ``Topology and data,'' \emph{Bulletin of the American Mathematical
  Society}, vol.~46, no.~2, pp. 255--308, 2009.

\bibitem{Ortega2018GraphSignalProcessing}
A.~{Ortega}, P.~{Frossard}, J.~{Kovačević}, J.~M.~F. {Moura}, and
  P.~{Vandergheynst}, ``Graph signal processing: Overview, challenges, and
  applications,'' \emph{Proceedings of the IEEE}, vol. 106, no.~5, pp.
  808--828, 2018.

\bibitem{takens1981detecting}
F.~Takens, ``Detecting strange attractors in turbulence,'' in \emph{Dynamical
  systems and turbulence, Warwick 1980}.\hskip 1em plus 0.5em minus 0.4em\relax
  Springer, 1981, pp. 366--381.

\bibitem{perea2015sliding}
J.~A. Perea and J.~Harer, ``Sliding windows and persistence: An application of
  topological methods to signal analysis,'' \emph{Foundations of Computational
  Mathematics}, vol.~15, no.~3, pp. 799--838, 2015.

\bibitem{robinson2016quasiperiodic}
M.~{Robinson}, ``A topological low-pass filter for quasi-periodic signals,''
  \emph{IEEE Signal Processing Letters}, vol.~23, no.~12, pp. 1771--1775, 2016.

\bibitem{trautmann1995physical}
S.~Trautmann, ``{A Physical String Model with a Twist},'' in \emph{Proceedings
  of the International Computer Music Conference (ICMC)}, 1995.

\bibitem{Essl:2005ph}
G.~Essl, ``{Aspects of the Topology of Interactions on Loop Dynamics in One and
  Two Dimensions},'' in \emph{LNCS 3310 Proceedings of the International
  Symposium on Computer Music Modeling and Retrieval 2004}.\hskip 1em plus
  0.5em minus 0.4em\relax Esbjerg, Denmark: Springer Verlag, 2005, pp.
  220--231.

\bibitem{marvasti2001nonuniform}
F.~Marvasti, \emph{Nonuniform sampling: theory and practice}.\hskip 1em plus
  0.5em minus 0.4em\relax Springer, 2001.

\bibitem{miskowicz2018event}
M.~Miskowicz, \emph{Event-based control and signal processing}.\hskip 1em plus
  0.5em minus 0.4em\relax CRC press, 2016.

\bibitem{gray1979fragments}
J.~W. Gray, ``Fragments of the history of sheaf theory,'' in \emph{Applications
  of sheaves}.\hskip 1em plus 0.5em minus 0.4em\relax Springer, 1979, pp.
  1--79.

\bibitem{shepard1986cellular}
A.~D. Shepard, ``A cellular description of the derived category of a stratified
  space,'' Ph.D. dissertation, Brown University, 1986.

\bibitem{curry2014sheaves}
J.~M. Curry, ``Sheaves, cosheaves and applications,'' Ph.D. dissertation, The
  University of Pennsylvania, 2014.

\bibitem{bredon1997sheaf}
G.~E. Bredon, \emph{Sheaf Theory, Graduate Texts in Mathematics}.\hskip 1em
  plus 0.5em minus 0.4em\relax Springer New York, 1997.

\bibitem{FILTERS07}
J.~O. Smith, \emph{Introduction to Digital Filters with Audio
  Applications}.\hskip 1em plus 0.5em minus 0.4em\relax W3K Publishing, 2007.

\bibitem{gallier2019algebra}
\BIBentryALTinterwordspacing
J.~Gallier and J.~Quaintance, \emph{Algebra, Topology, Differential Calculus,
  and Optimization Theory For Computer Science and Machine Learning}.\hskip 1em
  plus 0.5em minus 0.4em\relax self-published, 2019, book in Progress.
  \url{https://www.cis.upenn.edu/~jean/math-deep.pdf}. Retrieved on March 20,
  2020. [Online]. Available:
  \url{https://www.cis.upenn.edu/~jean/math-deep.pdf}
\BIBentrySTDinterwordspacing

\bibitem{robinson2015sheaf}
M.~Robinson, ``A sheaf-theoretic perspective on sampling,'' in \emph{Sampling
  Theory, a Renaissance}, G.~E. Pfander, Ed.\hskip 1em plus 0.5em minus
  0.4em\relax Springer, 2015, pp. 361--399.

\end{thebibliography}

\end{document}